# Tunable p-n junction barriers in few-electron bilayer graphene quantum dots

Fang-Ming Jing[1,3], Guo-Quan Qin[1,3], Zhuo-Zhi Zhang[1,2,3],

Xiang-Xiang Song[1,2,3,*], and Guo-Ping Guo[1,3,4]

1. CAS Key Laboratory of Quantum Information, University of Science and Technology of China, Hefei, Anhui 230026, China
2. Suzhou Institute for Advanced Research, University of Science and Technology of China, Suzhou, Jiangsu 215123, China
3. CAS Center for Excellence in Quantum Information and Quantum Physics, University of Science and Technology of China, Hefei, Anhui 230026, China
4. Origin Quantum Computing Company Limited, Hefei, Anhui 230088, China

* Author to whom correspondence should be addressed: songxx90@ustc.edu.cn

# Abstract

Graphene quantum dots provide promising platforms for hosting spin, valley, or spin-valley qubits. Taking advantage of the electrically generated band gap and the ambipolar nature, high-quality quantum dots can be defined in bilayer graphene using natural p-n junctions as tunnel barriers. In these devices, demonstrating the electrical tunability of the p-n junction barriers and understanding its physical mechanism, especially in the few-electron regime, are essential for further manipulating electron's quantum degrees of freedom to encode qubits. Here, we show the electrostatic confinement of single quantum dots in bilayer graphene using natural p-n junctions. When the device is operated in the few-electron regime, the electron tunneling rate is found to be monotonically tuned by varying gate voltages, which can be well understood from the view of manipulating the p-n junction barriers. Our results provide an insightful understanding of electrostatic confinement using natural p-n junctions in bilayer graphene, which is beneficial for realizing graphene-based qubits.

Graphene is considered a promising material for hosting spin qubits due to its weak spin-orbit coupling strength and hyperfine interaction.[1-3] Meanwhile, the highly tunable valley degree of freedom makes graphene attractive for encoding valley or spin-valley qubits as well.[4-6] To achieve these goals, gate-controlled quantum dots (QDs) based on graphene have been proposed and realized since they offer platforms for manipulating various quantum degrees of freedom at single-particle level.[7]

One of the key steps towards realizing qubits in QDs is controlling the tunnel barriers which define the dot. Although QDs have been studied for more than a decade in monolayer graphene (MLG),[8-11] early devices suffer from localized edge states induced in the etching process for shaping MLG flakes to QDs.[12,13] In these devices, tunnel barriers are usually difficult to be monotonically tuned by gate voltages,[14] making it challenging to operate the device in the few-electron regime. Recently, making use of bilayer graphene (BLG), in which a band gap can be electrically generated,[15,16] electrostatically confined QDs are achieved.[17-20] Different mechanisms are employed to form tunnel barriers. One is to use gate voltages to locally lift the Fermi level of BLG underneath the gate electrode into the band gap,[5,21,22] similar to QDs defined in traditional semiconductors (such as Si[23-25] and GaAs[26-29]) and other 2D semiconducting materials (such as $MoS_2$[30-32] and $WSe_2$[33-35]). In these devices, tunnel barriers exhibit high tunability,[36,37] which can be understood as similar to their counterparts formed in other semiconductors. The other way is to take advantage of the ambipolar nature of BLG to use natural p-n junctions as tunnel barriers, enabling defining QDs with fewer gate electrodes.[38-42] Systematic studies on these natural p-n junction barriers, especially in the few-electron regime, are of particular importance. Although tunable tunneling rates upon varying barrier gate voltages have been reported recently,[5] developing a flexible multi-gate tuning strategy of the natural p-n junction barriers and understanding its physical mechanism are highly desirable. This is because they are essential for manipulating electron's quantum degrees of freedom towards encoding qubits in graphene.

Here we demonstrate BLG QDs with natural p-n junctions as tunnel barriers. The device can be operated as a single dot, reaching the few-hole and the few-electron regimes, respectively. In the few-electron regime, we demonstrate that the rate at which electrons tunnel through the dot can be monotonically tuned by varying voltages applied to either the neighboring finger gates located near the dot, or the back gate sitting below the dot. The gate tuning effect is well-understood from the view of manipulating the natural p-n junction barriers. Our results are beneficial for confining and manipulating QDs in BLG using natural p-n junctions as tunnel barriers, which pave the way towards graphene-based quantum computation.

A schematic diagram of a typical device is illustrated in Fig. 1(a). The BLG flake is encapsulated by top and bottom hexagonal boron nitride (hBN) flakes using the van der Waals pick-up technique,[43,44] forming a sandwich-like heterostructure. After being placed on a pre-patterned metal back gate, the heterostructure is annealed at 450 °C in Ar/$H_2$ atmosphere for hours to remove residuals. Then reactive ion etching followed by evaporation of Cr/Au are employed to fabricate metal contacts.[43] Two split gates (Ti/Pd), with a designed separation of 100 nm in between, are deposited on the top hBN flake. After depositing an insulating layer of aluminum oxide, a set of finger gates (Ti/Pd) are fabricated. The cross-sectional schematics of fabrication process are shown in Fig. 1(b).

Figure. 1(c) shows an optical microscope image of the fabricated device. An atomic force microscope image that details the region of the transferred heterostructure on the back gate (corresponding to (i) in Fig. 1(b)) is shown in Fig. 1(d), suggesting clean interfaces. The designed metal contacts (blue), split gates (cyan) and finger gates (magenta) are kept away from bubbles in the heterostructure. The inset of Fig. 1(d) shows a scanning electron microscope image of the fabricated finger gates and the channel between the split gates. All electrodes are fabricated using standard e-beam lithography followed by e-beam evaporation. The device is measured in a dilution refrigerator with a base temperature of ~20 mK.

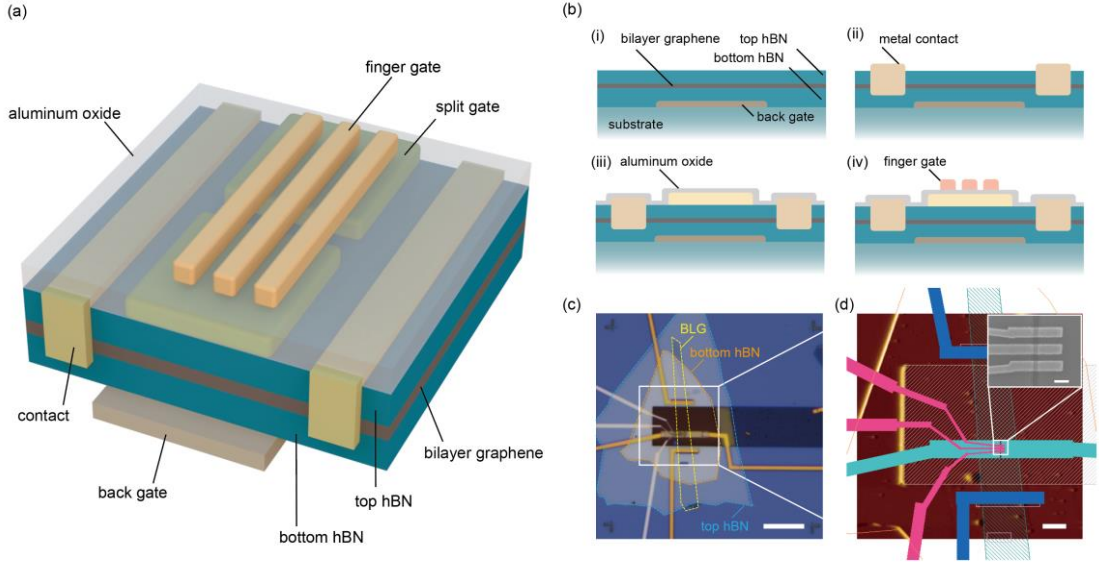

Figure 1. (a) Schematic of a gate-defined BLG QD device. (b) Cross-sectional schematics of device fabrication process. (c) Optical microscope image of a fabricated device with BLG and hBN flakes highlighted in different colors. The scale bar corresponds to 10 μm. (d) Atomic force microscope image of the region in the white box in (c). The image is taken after transferring the heterostructure onto the pre-patterned back gate. The designed metal contacts (blue), split gates (cyan) and finger gates (magenta) are kept away from bubbles. The scale bar corresponds to 2 μm. Inset: Scanning electron microscope image of the region where the QD is defined. The scale bar corresponds to 200 nm.

QDs are defined in the following way. First, voltages with opposite polarities are applied to the back gate and the split gates to create a perpendicular electric field to introduce a band gap in BLG. By properly adjusting the gate voltages, the BLG regions underneath the two split gates are tuned to be insulating, leaving a conducting channel in between. Then, multiple finger gates, which are above the channel, are responsible for manipulating the potential to define QDs. Taking an example, Figure. 2(a) shows the formation of a hole QD. The left panel shows the measured differential conductance $dI/dV$ as a function of the voltage $V_{FG2}$ applied to the second finger gate FG2. Here, an n-doped channel is formed while the regions below the split gates are gapped (back gate voltage $V_{BG} = 2$ V, split gate voltages $V_{SG1} = -2.37$ V and $V_{SG2} = -2.35$ V are applied). Three regimes can be found in the measured conductance curve: 1) When $V_{FG2} > -4$ V, the channel is overall of n-type. Therefore, a pronounced conductance can be measured. 2) When decreasing $V_{FG2}$, the conductance is suppressed, suggesting the Fermi level of BLG below FG2 is gradually tuned into the band gap. 3) When $V_{FG2} < -5$ V, a p-type island is locally formed underneath FG2. The natural junctions between p-type and n-type regions act as tunnel barriers, defining a QD filled with holes. As a consequence, a series of differential conductance peaks are found, suggesting the filling of the first four holes. The corresponding band profile schematic is depicted in the right panel of Fig. 2(a).

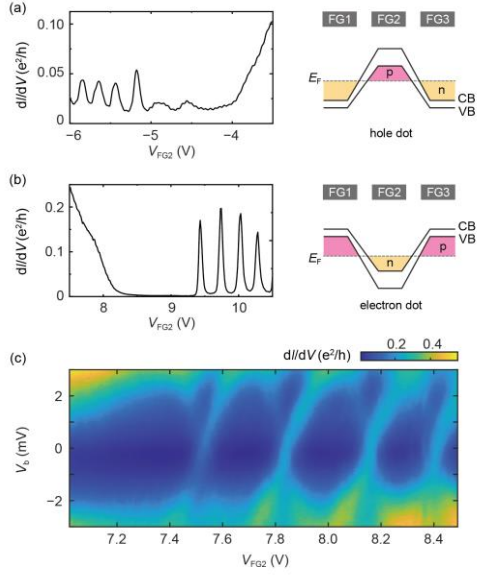

Figure 2. (a) Left panel: Measured differential conductance as a function of the voltage $V_{FG2}$ applied to the second finger gate FG2. Right panel: Corresponding band profile schematic in the conducting channel defined by the split gates, showing the formation of a hole QD with natural p-n junctions as tunnel barriers. FG1, FG2, and FG3 represent the three finger gates that are placed above the channel. The Fermi level $E_F$ with respect to the conduction band (CB) and valence band (VB) results in corresponding p-type and n-type conducting regimes. (b) Measured differential conductance as a function of $V_{FG2}$ (left panel) and the corresponding band profile schematic (right panel), showing the formation of an electron QD. (c) First four Coulomb diamonds measured from an electron QD. Here $V_{BG} = -3$ V, $V_{SG1} = 1.92$ V, $V_{SG2} = 1.89$ V, $V_{FG1} = -3$ V, $V_{FG3} = 0$ V.

Taking advantage of the ambipolar nature of BLG, the QD can be filled with electrons alternatively, when applying opposite voltages (for example, $V_{BG} = -3.75$ V, $V_{SG1} = 2.59$ V, $V_{SG2} = 2.55$ V). As shown in Fig. 2(b), similarly, an electron QD that employs natural p-n junctions as tunnel barriers is realized. Note that for the electron dot formation, the differential conductance can be completely pinched off at the blockade regime, compared with that of the hole dot formation. This is possibly due to disorder-induced hopping paths remained at the gate voltage configuration for defining a hole dot. Below, we focus on studying the electron dot. Figure 2(c) shows a finite bias spectroscopy of the electron dot. A finite bias voltage $V_b$ is applied between the source and the drain, and the differential conductance as a function of $V_{FG2}$ is measured. Diamond-shaped regions of Coulomb blockade are distinguished, within which the electron number in the dot is fixed.[27] As demonstrated in Fig. 2, the highly tunable BLG QD can be operated in either the few-hole or the few-electron regime. In both regimes, the dot can be completely depleted.

Next, we investigate the tunability of the p-n junction barriers in the few-electron regime. Figure 3(a) shows the first four Coulomb peaks (marked using different labels) at $V_{BG} = -3.0$ V, as a function of $V_{FG2}$ and voltages simultaneously applied to the neighboring finger gates ($V_{FG1}$&$V_{FG3}$). The conductance peaks are parallel to each other, suggesting the formation of a single dot.[26] The heights of the first four Coulomb peaks are monotonically suppressed when increasing $V_{FG1}$&$V_{FG3}$, which is highlighted in Fig. 3(b). This suggests that the overall tunneling rate $\Gamma$ is also monotonically decreased since the conductance can be expressed as $e\Gamma/V_b$, where $e$ is the electron charge.[45,46] The effect of tuning $V_{BG}$ on the natural p-n junction barriers is also studied. Figure 3(d) shows the first four Coulomb peaks as a function of finger gate voltages measured at $V_{BG} = -4.5$ V, suggesting a single dot formation as well. Similarly, suppression of conductance upon increasing $V_{FG1}$&$V_{FG3}$ is observed. Meanwhile, we find for $V_{BG} = -4.5$ V, the peak conductance is smaller than that at $V_{BG} = -3.0$ V. As shown in Fig. 3(e), we plot the heights of the first four conductance peaks as a function of $V_{BG}$ for fixed values of $V_{FG1}$ and $V_{FG3}$. For both cases of $V_{FG1} = V_{FG3} = -2.0$ V and $V_{FG1} = V_{FG3} = -0.5$ V, similar trends of conductance increase when increasing $V_{BG}$ are found.

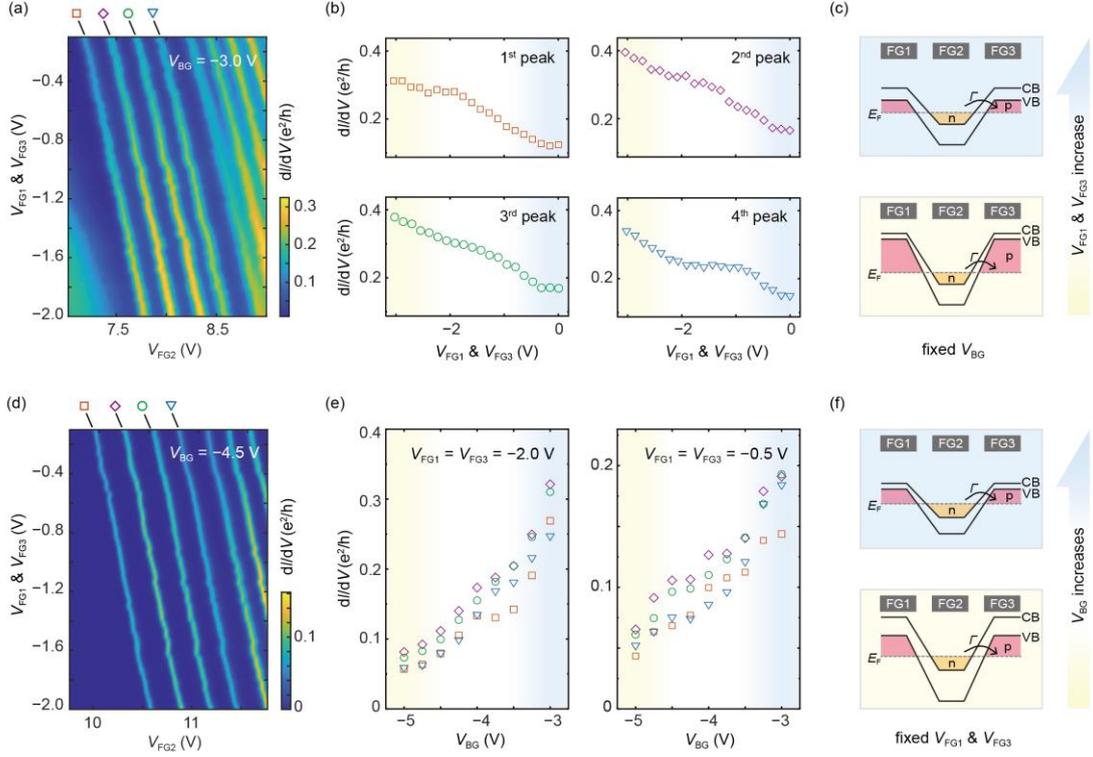

Figure 3. (a) Differential conductance as a function of $V_{FG2}$ and voltages ($V_{FG1}$&$V_{FG3}$) simultaneously applied to the neighboring finger gates (FG1 and FG3). Here $V_{BG} = -3.0$ V, $V_{SG1} = 1.92$ V, $V_{SG2} = 1.89$ V. (b) Peak heights of the first four conductance peaks, marked by the corresponding labels in (a), as a function of $V_{FG1}$&$V_{FG3}$ at $V_{BG} = -3.0$ V. (c) Band profile schematics illustrating the effect of tuning finger gate voltages $V_{FG1}$ and $V_{FG3}$ on the tunneling rate $\Gamma$. (d) Differential conductance as a function of $V_{FG2}$ and $V_{FG1}$&$V_{FG3}$. Here $V_{BG} = -4.5$ V, $V_{SG1} = 3.23$ V, $V_{SG2} = 3.19$ V. (e) Peak heights of the first four conductance peaks, marked by the corresponding labels in (a) and (d), as a function of $V_{BG}$ for fixed $V_{FG1}$&$V_{FG3}$ values of $-2.0$ V and $-0.5$ V, respectively. (f) Band profile schematics illustrating the effect of tuning back gate voltage $V_{BG}$ on the tunneling rate $\Gamma$.

To understand the physical mechanism of the gate tuning effect on the tunneling rate, we schematically depict the evolution of the band profile upon varying gate voltages, as illustrated in Figs. 3(c) and 3(f). It is worth noting that within the applied gate voltage ranges ($V_{FG1}$, $V_{FG3}$, and $V_{BG}$), variations in the size, as well as the position, of the dot can be neglected. This is revealed by analyzing peak spacings between Coulomb peaks and slopes of Coulomb peaks in the space of $V_{FG2}$ and $V_{FG1}$&$V_{FG3}$ (see Supplementary Material). This behavior is in sharp contrast to the previous work, where the dot size changes by a factor of ~3 when varying gate voltages.[5] Since variations in the dot size can modify the sharpness of the p-n junctions thus tunnel barriers, it is beneficial to keep the dot size unchanged to better illustrate the gating effect on the band profile.

Figure 3(c) shows the band profile schematics of tuning $V_{FG1}$ and $V_{FG3}$ at fixed $V_{BG}$. When $V_{BG}$ is fixed at $-3$ V, increasing $V_{FG1}$ ($V_{FG3}$) from $-3$ V to $0$ V makes the local perpendicular electric field between FG1 (FG3) and the back gate stronger, thus enlarging the band gap of the BLG underneath FG1 (FG3). At the meantime, increasing $V_{FG1}$ ($V_{FG3}$) decreases the hole density below FG1 (FG3). The combined contribution results in thicker p-n junction tunnel barriers, leading to a decrease of measured conductance.

The effect of tuning the tunneling rate upon changing $V_{BG}$ at fixed $V_{FG1}$ and $V_{FG3}$ can be understood from the view of manipulating the p-n junction barriers as well. Taking the example of $V_{FG1} = V_{FG3} = -2.0$ V, when $V_{BG}$ is increased from $-5$ V to $-3$ V, the band gap of BLG in the whole channel is decreased. Meanwhile, the Fermi level is lifted, resulting in a decrease of the hole density underneath FG1 and FG3. The combined effect results in narrower p-n junction tunnel barriers, thus larger conductance measured at larger $V_{BG}$. The corresponding band profile schematics of tuning $V_{BG}$ are illustrated in Fig. 3(f). Note that increasing $V_{BG}$ also increases the electron density underneath FG2. Therefore, much smaller $V_{FG2}$ is needed to balance this influence to maintain a similar size of the dot at larger $V_{BG}$. This explains the different $V_{FG2}$ ranges between Figs. 3(a) and 3(d).

In conclusion, using hBN encapsulation and multilayer gate electrodes, we demonstrate the formation of a single QD that can be operated in the few-electron regime. The tunnel barriers for defining the dot are formed by natural p-n junctions in the conducting channel. The heights of the first four Coulomb peaks are monotonically tuned upon varying $V_{FG1}$, $V_{FG3}$ and $V_{BG}$, which can be well-understood from the view of tunable p-n junction barriers. We would like to point out that based on the device with these tunable natural p-n junction barriers, confined spin and valley states in BLG can be manipulated at single-particle level,[17,47] offering the possibility of encoding qubits. Moreover, the demonstrated tunable natural p-n junction barriers are applicable to defining double, and triple dots in BLG,[36,37,48-50] allowing further qubit integrations based on multiple QDs. Our results deepen the understanding of electrostatic confinement in BLG using natural p-n junctions, offering promising future of realizing spin, valley, and spin-valley qubits based on BLG QDs.

## Supplementary Material

Analysis of peak spacings and slopes of Coulomb peaks can be found in the Supplementary Material.

## Acknowledgments

This work was supported by the National Natural Science Foundation of China (Grant Nos. 12274397, 11904351, 12274401, and 12034018). This work was partially carried out at the USTC Center for Micro and Nanoscale Research and Fabrication.

## Conflict of Interest

The authors have no conflicts to disclose.

## Data Availability

The data that support the findings of this study are available from the corresponding authors upon reasonable request.